\documentclass[12pt]{article}
\def\lromn#1{\uppercase\expandafter{\romannumeral#1}}

\usepackage{graphicx}

\begin{document}
\begin{flushright}
KOBE-TH-04-06 
\end{flushright}
\begin{center}
\begin{large}

\vspace*{1cm} 

\textbf{
A variety of lepton number violating processes related to 
Majorana neutrino masses
}

\end{large}

\vspace{1cm}
\begin{large}
C.S. Lim$^{1}$, E. Takasugi$^{2}$ and M. Yoshimura$^{3}$

\vspace*{0.5cm}

$^{1}$
Department of Physics, Kobe University \\ 
Rokkodai 1-1, Kobe 657-8501, Japan  

$^{2}$
Department of Physics, Osaka University \\ 
Toyonaka, Osaka 560-0043, Japan 

$^{3}$
Department of Physics, Okayama University \\
Tsushima-naka 3-1-1, Okayama 700-8530, Japan 
\end{large}

\vspace{1.5cm}

{\bf ABSTRACT}
\end{center}

A Majorana type of the neutrino mass matrix induces a class of lepton number
violating processes.
Cross sections of these reactions are given in terms of the 
neutrino mass matrix element, and a semi-realistic event rate is estimated.
These processes provide mass and mixing parameters not directly
accessible by the neutrino oscillation experiments.
If these processes are discovered with a larger rate than given here,
it would imply a new physics of the lepton number violation not directly
related to the Majorana neutrino mass, such as R-parity violating operators
in SUSY models.

\newpage

Neutrino oscillation observed in SuperKamiokande \cite{sk}, SNO \cite{sno}, 
and KamLAND \cite{kam}
experiments have opened a new window beyond physics of the standard model. 
The immediate critical question is to determine the nature of neutrino masses;
whether they are of Dirac or of Majorana type. 
Unfortunately, the data on the neutrino oscillations can not discriminate 
between the types of neutrino masses, as we will see below. 
In order to get a conclusive argument on this issue, the investigation into 
lepton number violating processes, inevitable consequence of the Majorana nature, 
is indispensable. 

If the Majorana mass is verified, it opens up the possibility of
generating the baryon asymmetry of the universe via leptogenesis
scenario \cite{fy}.
A conventional lepton number violating process and the one most extensively discussed towards
this goal is the neutrinoless double beta decay \cite{2beta}.
Despite several ingenious experimental proposals for improved detection 
of this decay
we believe that some alternative methods to measure the Majorana type
of the neutrino masse matrix are both useful and very important.
In the present work 
we examine a variety of lepton number violating processes for 
this purpose.

We systematically examine 
a class of lepton number violating effective operators below
the Fermi energy scale $1/\sqrt{G_{F}}$, 
\begin{equation}
ll\, \bar{q}q\, \bar{q}q \; ,
\end{equation}
where $l$ is the lepton doublet and $q$ is the quark doublet having
the quantum number of the standard model.
Existence of this class of operators requires a new physics
beyond the standard theory, but we do not need to specify the
new physics.
The Feynman diagram 
that generates this class of effective operators is
depicted in Figure 1; the cross in the figure is the Majorana neutrino
mass matrix and the exchanged particle is a weak boson, either $W^{\pm}$
or $Z$.  
\begin{figure}[ht]
\begin{center}
\includegraphics[width=8cm]{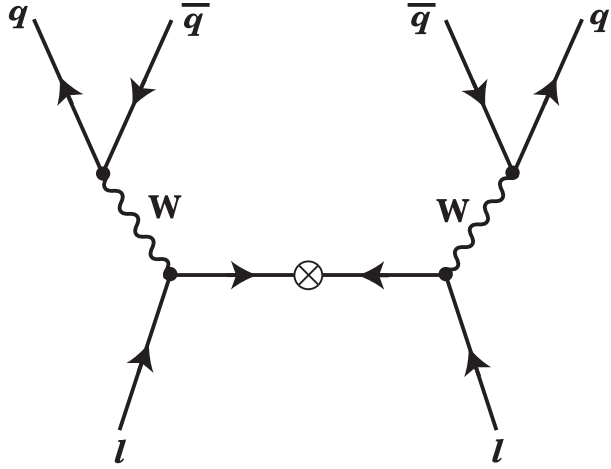}
\end{center}
\caption{\label{fig1}The Feynman diagram 
that generates effective lepton number violating operators}
\end{figure} 
When both of $l$ are the electron and $q$ is either a u- or d-type
quark, the operator gives the neutrinoless double beta decay.
We extend this double beta process to all combination of charged
leptons of $l$, such that the full neutrino mass matrix element may be
explored.

The strength of these dimension 9 operators at low energy scale
of $\sqrt{s} \leq 250 GeV$ is of order
\begin{equation}
G_F^2 m_\nu s^{3/2} \sim 10^{-22} \frac{m_{\nu}}{1eV}\, 
(\frac{\sqrt{s}}{100 MeV})^3 \; .
\end{equation}
This is an extremely small number compared to the usual weak interaction
of order $G_F s$, but the nature of neutrino mass may 
well be examined only by these operators.
The important point is that once the Majorana nature of the neutrino mass is
assumed, there is no arbitrary freedom left, both for existence and
its strength of the class of lepton number violating processes 
we consider below. 

Cross sections of processes discussed below are
determined by using the matrix element of the Majorana neutrino mass,
\begin{equation}
m_{\alpha \, \beta} = \Sigma_k U_{\alpha k}U_{\beta k}\,m_k \;,
\end{equation}
where $(k = 1,2,3)$ and $(\alpha \,, \beta = e, \mu, \tau)$
and $m_k$'s are the mass eigenvalues (real and positive by definition).
The reaction rate is 
thus independent of the origin of the neutrino mass matrix such as
the seesaw mechanism \cite{seesaw}.
Note that the above combination of neutrino masses and the mixing
parameters is different from that measured in the neutrino oscillation
experiment, in which $ U_{\alpha k}$
and its conjugate $U_{\beta k}^*$ appears, 
thus the Majorana phases, characteristic of the Majorana nature, just cancel 
out in neutrino oscillation. 
Since the Majorana phase factor plays an important role in leptogenesis calculation, in general, the investigation into the lepton number violating
reactions is relevant for the leptogenesis.

If one of these processes is discovered with a larger rate
than given below,
it means existence of a new class of diagrams not involving
the neutrino mass matrix and may provide a new feature to the lepton
sector such as R-parity violating interactions in SUSY models.
In that sense lepton number violating processes 
complementary to the neutrinoless double beta decay 
may provide the unique window to determine the mechanism of
how the lepton number violation occurs.

With the lepton flavor mixing, the effective operators above give a variety
of lepton number violating processes. We consider the following reactions
which we think relatively easy to explore experimentally;
\begin{eqnarray}
&&
(e \mu); \hspace{1cm}
e^- + A^Z \rightarrow \mu^+ + A^{Z-2} \;,
\label{e-mu} 
\\ &&
(\mu e); \hspace{1cm}
\mu^-  + A^Z \rightarrow e^+ + A^{Z-2} \;,
\label{mu-e}
\\ &&
(e \mu); \hspace{1cm}
p + A^Z \rightarrow \mu^+ +e^+ + (A+1)^{Z-1} \;, 
\label{pa}
\\ &&
(e e); \hspace{1cm}
e^- + e^- (atomic) \rightarrow \pi^- + \pi^- \;,
\label{ee}
\\ &&
(e \mu); \hspace{1cm}
\nu_{\mu} + e^- (atomic) \rightarrow \pi^- + \pi^0 \; .
\label{e-nu}
\end{eqnarray}
Here $(\alpha \beta)$ means that the process may explore the matrix element
$m_{\alpha \, \beta}$.
There are corresponding inclusive processes such as
\(\:
e^- + A^Z \rightarrow \mu^+ + X \;,
\:\)
where $X$ is any hadronic state.
When $A$ is a light nucleus, it may break up due to a low
binding energy, for instance,
\begin{equation}
e^- + He^4 \rightarrow \mu^+ + 4\,n \;.
\end{equation}

To reduce the background at low energies to a respectable level, 
it is important to search for unambiguous signatures , for instance
without decay muons
below the $\pi$ production threshold for the first process of (\ref{e-mu}).
On the other hand, at high enough energies pions may not decay within the
detector volume, in which case one does not worry about $\mu^{\pm}$ 
from the pion decay.
It is also important to use a high intensity beam and a high density
target in order to overcome the low reaction rate, 
for instance the atomic electron target
in the processes of (\ref{ee}) and (\ref{e-nu}).
We also consider muons captured by nucleus in (\ref{mu-e}) to compensate
for the low flux.
The other processes we considered and not listed here
are difficult due to various reasons;
the background  problem, the insufficient beam energy 
available or planned now, a finite short lifetime and so on.

We consider important  processes in turn, and
the conversion processes, 
$e^- + He^4 \rightarrow \mu^+ + 4\,n$ 
and 
$e^- + A^Z \rightarrow \mu^+ + A^{Z-2}$
are the first.
A related conversion process,
\(\:
\mu^- + A^Z \rightarrow e^+ + A^{(Z -2)}
\:\),
\(\:
\mu^- + A^Z \rightarrow \mu^+ + A^{(Z -2)}
\:\),
has been considered in \cite{2beta} and \cite{mmm}.
The experiment could be done effectively below the $\pi$-production
threshold rejecting the $\mu^+$ background from the $\pi^+$ decay, 
hence for this case the incident electron energy should be chosen properly.
For the 2nd process of nuclear target, 
the cross section at low energies is given by
\begin{equation}
\sigma=\sum_{f}\sigma_{i\to f} = \frac{G_F^4 m_{e \mu}^{2} <p_+>< E_+>}
{32 \pi^3 R_n^2}\sum_{f,s}|t^{\mu \nu}\Omega_{\mu\nu}|^2\;,
\end{equation}
where $t^{\mu\nu}$ is the lepton wave function, $\Omega_{\mu\nu}$ is 
the nuclear matrix element and $R_n$ is an effective nuclear radius, 
as defined in \cite{double beta}. 
Since the final nucleus can take any final 
state, we sum up the final states. Therefore, we use the average momentum 
and energy for $\mu^+$. We neglect the momentum of $\mu$ in $t^{\mu\nu}$ 
and use the approximation to evaluate the nuclear matrix element 
\cite{double beta}. 
We find $\sum_{f,s}|t^{\mu \nu}\Omega_{\mu\nu}|^2\simeq$$ 
(g_V^2+18g_A^2+9g_A^4)Z(Z-1)$ for $S=0$ and 
$(g_V^2+2g_A^2+9g_A^4)Z(Z-1)$ for $S=1$. Here, 
$S$ is the spin of two protons and $Z$ is the atomic number. 
From this, we find with $Z=50$ and $A=100$
\begin{equation}
\sigma \sim 5\times  10^{-65}\, cm^2\,(\frac{|m_{e \mu}|}{100 eV})^2
\;,
\end{equation}
for the average $\mu$ momentum of $30 MeV$. It may be worthwhile to 
comment that there is  no reduction due to the nuclear matrix element
in comparison with the double beta decay and also there is 
enhancement due to many possible combinations of the proton target 
inside the nucleus.

The helium case can roughly be estimated by just taking $Z=2$, and
we find 
\begin{equation}
\sigma \sim  2\times 10^{-67}\, cm^2\,(\frac{|m_{e \mu}|}{100 eV})^2
\;,
\end{equation}
In order to compensate for the small reaction rate at low energies,
it might be useful to go to higher energies despite larger
backgrounds. Energy dependence of the cross section goes like
$\propto s^2$ with $\sqrt{s}$ the CMS energy and it is 
\(\:
O[1]\,G_F^4|m_{ e \mu}|^2\,s^2/16 \pi^3
\:\).
A simple parton model neglecting transverse momentum may be used in estimating
$O[1]$ factor here.

The advantage of muon capture listed in (\ref{mu-e}) is two fold;
it automatically gives a self-focusing mechanism of the incident
particle, capture into the area
of nuclear size, and the same muon may repeatedly be used 
as in the case of a high luminosity accumulator ring.
Thus, a naive computation yields a very large event rate, for instance
even with
a pulsed muon flux of order $10^{12} /sec$ much smaller than the highest
achieved electron flux.
The problem however is that at the same time the first order weak process,
namely muon capture of $\mu^{-} \rightarrow \nu_{\mu}$ also
becomes huge, and there is no chance of survival of captured muons
left for the lepton number violating processes.
Technically, the relevant quantity in this case is the branching fraction
of muon captured reactions instead of the absolute event rate, 
and for this quantity the enhancement factor
due to the capture into the atomic orbit cancels.
For example, the branching fraction of the process $\mu^- \rightarrow e^+$
is given by $O[G_F ^2 |m_{\mu e}|^2Z/8\pi^2 R^2]\sim 3\times 10^{-29}
|m_{\mu e}/100{\rm eV}|^2$[5].

We next discuss the  processes off atomic electrons,
(\ref{ee}) and (\ref{e-nu}).
Both processes are quite unique in that the lepton number
disappears in the final state. 
For the atomic target the threshold for both processes opens up at around
\(\:
\frac{2m_{\pi}^2}{m_e} \approx
80 GeV
\:\). 
The cross section for the process $e^- + e^-$ is given by
\begin{equation}
\sigma = \frac{G_F^4 f_{\pi}^4 |m_{e e}|^{2}}{2\pi}\,
\sqrt{\frac{s - 4m_{-}^2}{s}} \;,
\label{ee-sigma}
\end{equation}
where $f_{\pi}$ is the $\pi$ decay constant of order $90 MeV$,
and $s \approx 2m_e E_e$ is the CMS energy squared.
We have treated the atomic electron being at rest, since its momentum
is much smaller than the incident beam energy $E_e (\gg m_e)$.
The important prefactor 
\(\:
\frac{G_F^4 f_{\pi}^4 |m_{e \mu}|^{2}}{2\pi} 
\:\)is numerically
\(\:
\approx 
8 \times 10^{-67}\,cm^2\,(|m_{e \mu}|/100 eV)^2
\:\).
The event rate assuming an effective flux of order 
$10^{34}cm^{- 2}sec^{- 1}$ and a heavy target of 
mass $ 500 g$ and $A \sim 2Z$ is 
\(\:
\approx 4 \times 10^{2} \,(|m_{e \mu}|/100 eV)^2/ year
\:\).
A non-trivial background of the usual electromagnetic origin is
$e^- + n \rightarrow e^- + p + \pi^- + \pi^- + \pi^+$ 
with a missing $p$.
The detector should be arranged hermetically not to miss particles 
produced in the final state, and the kinematical condition
of the two-body process of (\ref{ee}) should be maximally
exploited to reject these backgrounds.

The corresponding inclusive process,
\(\:
e^- + e^- \rightarrow X 
\:\),
with $X$ any hadronic state
has a larger cross section of order,
\(\:
10^{-4}\,G_F^4|m_{ e e}|^2\,s^2
\:\),
or \\
\(\:
\approx 10^{-58}\,cm^2\,(|m_{e e}|/100 eV)^2\,(\sqrt{s}/100 GeV)^4
\:\)
for 
\(\:
\sqrt{s} \leq 250 GeV
\:\).
A possible experimental problem would be photons from $\pi^0$ decay
which might be misidentified by electrons at high energies. Thus, it would be
preferable, but difficult, to solely detect all charged modes 
not containing $\pi^0$ in the final state.
Furthermore, for the atomic electron target a reasonable choice of
the maximal $\sqrt{s}$ 
would be of order $1 GeV$, and even for this one needs a
TeV electron LINAC.
A more precise formula of the inclusive cross section may be
computed by using the simple quark model of 3 colors, to give at
\(\:
2 m_{W} > \sqrt{s} \gg 2 m_{\pi}
\:\)
\begin{eqnarray}
&&
\sigma = \frac{G_F^4|m_{ee}|^2\,s^2}{4 \pi^5}
\,\int_0^1 dx_1 \int_0^{1-x_1} dx_2\,f(x_1 \,, x_2) 
\end{eqnarray}
where $\sqrt{s} x_i$ is the invariant mass of the quark pair, $\bar{u}_i d_i$.
The integral of the invariant mass distribution $f(x_1 \,, x_2)$ is about
0.07.

The neutrino beam is attractive from two reasons: the low background and
a high intensity $\nu_{\mu}$ beam may become available in the proposed
neutrino factory. 
The high intensity $\nu_{\mu}$ beam of energy of order $80 GeV$
may also be of interest due to the possibility
of the $\tau$-neutrino appearance experiment.
A related tri-muon process to measure
$m_{\mu \mu}  $ has also been considered in \cite{tri-muon},
which belongs to another class of effective operators,
\(\:
ll \bar{l}l \bar{q}q
\:\).
Incidentally, we considered this class of operators along with 
yet another class of
\(\:
ll \bar{l}l \bar{l}l
\:\),
searching all promising processes of the lepton number
violation.
Due to the experimental difficulty of determining 
the sign of the lepton number 
of neutrinos the process of \cite{tri-muon}, namely
\(\:
\nu_{\mu} + p \rightarrow \mu^+ + \mu^+ + \mu^- + X
\:\),
seems the only feasible process of both of these two classes
for signatures of unambiguous lepton number violation.

The cross section for (\ref{e-nu}) is just $1/4$ times of the rate
eq.(\ref{ee-sigma}), neglecting the pion mass difference.
A non-trivial background comes from the usual neutral current weak process,
$\nu_{\mu} + n \rightarrow \nu_{\mu} + p + \pi^- + \pi^0$ with a
missing $p$.
The true signal is kinematically selected by, for instance, plotting
the invariant mass($(p_- + p_0)^2$) distribution of two pions
(low energy $\pi^0$ may also help by requiring two photon showers of
$\pi^0 \rightarrow \gamma \gamma$ decay), which
peaks around at $\sqrt{2m_e E_{\nu}}$ for a given incident neutrino
energy $E_{\nu}$ .

Finally, a low energy proton beam may be used in (\ref{pa}).
The cross section at low energies for this process is 
$O[10^{-70}] cm^2$ for $Z = 50$.

A meaningful limit of the absolute scale of neutrino masses
is set by the current negative search of the neutrinoless
double beta decay, which gives a limit \cite{double beta limit},
for the combination,
\begin{equation}
|m_{e e}| = |\Sigma_k U_{ e k}^2\, m_k|  < 0.3 eV \; .
\end{equation}
This is a combination of the mass matrix element different from
that appearing in the above process of (\ref{e-mu}), (\ref{pa}), (\ref{e-nu}), 
namely $m_{e \mu}$, but is identical to that of (\ref{ee}).
For instance, with a phase cancellation the neutrinoless double beta decay
and the related process (\ref{ee}) might be suppressed, 
but the other processes involving $m_{e \mu}$ might be relatively strong.
On the other hand, WMAP has recently derived from a detailed fluctuation
map of the microwave background the following limit \cite{wmap}, 
\cite{mass limit} 
\(\:
m_{max} < 0.2 eV \,.
\:\)
If this limit is to be taken at a face value, it may not be necessary
to set a neutrino mass limit of order $1 eV$ or larger.
Nonetheless, we believe that terrestrial experiments are to be
supplemented to exclude with certainty even this range of neutrino masses.
Complementary and redundant information on neutrino mass matrix elements 
$m_{\alpha \, \beta}$ is crucial for further understanding of the lepton, 
hence the GUT sector.
Thus, despite very small rates experimental search for these new reactions
is very welcome.

A precise relation between the parameters of the neutrino oscillation
and the lepton number violating process is given by, 
\begin{equation}
\label{relation}  
\Sigma_{\gamma} m_{\alpha \gamma} m_{\beta \gamma}^{*}
 = \Sigma_k U_{\alpha k}U_{\beta k}^* \,m_k^2. 
\end{equation} 
The quantity in the right hand side can be determined 
by a precision data of the spectral distortion
of K2K and Kamland oscillation experiments, when it is
collaborated with  the absolute mass scale determined by
a direct mass measurement.
The (absolute values of) $m_{\alpha \beta}$ in the left hand side, on the other hand, could be measured 
through the lepton number violating processes. 
In the physical observable in Eq.(\ref{relation}), however, the information of Majorana 
phases is lost, due to the product of mass matrix $m$ and its hermitian conjugate $m^{\dagger}$. This is why 
we should study the lepton number violating processes, which are caused by the mass matrix 
$m$ itself containing the information of the CP violating Majorana phases. 

To clarify this we write the unitary matrix $U$ as 
\begin{equation} 
\label{unitary} 
U = U_{MNS}  P, 
\end{equation}  
where $U_{MNS}$ is the Maki-Nakagawa-Sakata matrix with 1 CP violating phase and 
$P = \mbox{diag} (1, e^{i \alpha}, e^{i \beta})$, with $\alpha$ and $\beta$ being 
CP violating Majorana phases. We easily find that the observable in (\ref{relation}) 
can be written as \\ 
$(U_{MNS} \cdot \mbox{diag} (m_{1}^{2}, m_{2}^{2}, m_{3}^{2}) \cdot 
U^{\dagger}_{MNS})_{\alpha \beta}$, where 
the matrix $P$ disappears.   
On the other hand, from Eq.(3) we realize that $m_{\alpha \beta}$ itself does depend on the Majorana 
phases: 
\begin{equation} 
\label{Massmatrix}
m_{\alpha \beta} = (U_{MNS} \cdot \mbox{diag} (m_{1}, m_{2}e^{2i\alpha}, m_{3}e^{2i\beta}) \cdot 
U^{t}_{MNS})_{\alpha \beta}. 
\end{equation}  

Having valuable information on $U_{MNS}$ and the mass eigenvalues $m_{1}, m_{2}, m_{3}$ from  
the neutrino oscillation experiments, by use of Eq.(\ref{Massmatrix}) we can evaluate each of 
$|m_{\alpha \beta}|$ and therefore the feasibility of lepton number violating processes. 
In the approximation of neglecting $\theta_{13}$ and taking the maximal mixing 
for the atmospheric neutrino oscillation, the unitary matrix $U_{MNS}$ is expressed by 
\begin{eqnarray}
U_{MNS} = \pmatrix{c_\odot&s_\odot&o\cr 
-s_\odot/\sqrt 2&c_\odot/\sqrt 2&-1/\sqrt 2\cr
-s_\odot/\sqrt 2&c_\odot/\sqrt 2&1/\sqrt 2\cr}
,
\end{eqnarray}
where $s_\odot =\sin \theta_\odot$ and $c_\odot=\cos \theta_\odot$. 
Substituting this expression for $U_{MNS}$ in Eq.(\ref{Massmatrix}), we can 
readily compute $|m_{\alpha\beta}|$ . 
We generally find that $|m_{\mu\mu}|=|m_{\tau\tau}|$ and $|m_{e\mu}|=|m_{e\tau}|$. 
To get further results, we consider three typical cases of neutrino mass eigenvalues, 
i.e., the hierarchical (H), the inverted hierarchical (IH) 
and the quasi-degenerate (QD) cases. For the case H, utilizing $m_3 
\simeq \sqrt{\Delta m_{atm}^2}>>m_2 \simeq \sqrt{\Delta m_{sol}^2}
>>m_1$,  
we find 
\begin{eqnarray}
|m_{ee}|&\simeq& s_\odot^{2} \sqrt{\Delta m_{sol}^2}\;, 
\nonumber\\
|m_{e\mu}|&\simeq& \frac{|s_{2\odot}|}{2\sqrt 2} \sqrt{\Delta m_{sol}^2}\;, 
\nonumber\\
|m_{\mu\mu}|&\simeq& \frac12 \sqrt{\Delta m_{atm}^2}
+ \frac{1}{2} c_\odot^2  c_{2(\alpha-\beta)}\sqrt{\Delta m_{sol}^2}\;,
\nonumber\\
|m_{\mu\tau}|&\simeq& \frac12 \sqrt{\Delta m_{atm}^2}
-\frac{1}{2} c_\odot^2  c_{2(\alpha-\beta)}\sqrt{\Delta m_{sol}^2}\;. 
\end{eqnarray}
Therefore, the $m_{ee}$ which is the effective mass for the neutrinoless 
double beta decay is small and also the contribution of the Majorana phases 
is suppressed. 
For the IH case, by using $m_1\simeq m_2\sim \sqrt{\Delta m_{atm}^2}>>m_3$, we 
find 
\begin{eqnarray}
|m_{ee}|&\simeq& m_1 \sqrt{1-s_{2\odot}^2s_\alpha^2} 
\ge m_1 |c_{2\odot}|\simeq \frac12 \sqrt{\Delta m_{atm}^2} \sim 0.03{\rm eV}\;,
\nonumber\\
|m_{e\mu}|&\simeq &\frac{|s_{2\odot}s_\alpha|}{\sqrt 2} \sqrt{\Delta m_{atm}^2}
\;,
\nonumber\\ 
|m_{\mu\mu}|&\simeq& |m_{\mu\tau}| \simeq \frac12 |m_{ee}|\;. 
\end{eqnarray}
In this case, the effective mass of neutrinoless double beta decay is bounded 
from below and the decay will be measurable in high precision experiments, thus 
providing a useful information on the Majorana phase $\alpha$. 
For the QD case, by using $m_1\simeq m_2 \simeq m_3 >\sqrt{\Delta m_{atm}^2}$, 
we find 
\begin{eqnarray}
|m_{ee}|&\simeq& m_1 \sqrt{1-s_{2\odot}^2s_\alpha^2}>0.03{\rm eV}\;,
 \nonumber\\
|m_{e\mu}|&\simeq &\frac{|s_{2\odot}s_\alpha|}{\sqrt 2} m_1\;,
 \nonumber\\
|m_{\mu\mu}|&\simeq& m_1\sqrt{1-(s_\odot^2c_\odot^2s_\alpha
^2+s_\odot^2s_\beta^2+c_\odot^2s_{\alpha-\beta}^2)}\;,
\nonumber\\
|m_{\mu\tau}|&\simeq& m_1\sqrt{1-(s_\odot^2c_\odot^2s_\alpha
^2+s_\odot^2c_\beta^2+c_\odot^2c_{\alpha-\beta}^2)}\;. 
\end{eqnarray}
Therefore, again we have an enough chance to measure the effective 
mass of the neutrinoless double beta decay. In addition, from the experiments to measure the 
other matrix elements we may be able to obtain the information on both of 
two Majorana phases $\alpha$ and $\beta$.

We thus learn that if the neutrinoless double beta decay unfortunately 
ends up with null result of precision of$O[0.03]$eV, only 
the normal hierarchical mass pattern is allowed. 
In this case the lepton number violation
can be verified in terrestrial laboratories only 
by lepton number violating processes discussed in the present work.

The interesting question is how CP violation, relevant for the leptogenesis, 
may be examined by studying the CP violating phases in the Majorama mass 
matrix of left-handed neutrinos, which are obtainable from the neutrino oscillation experiment, handled by $U_{MNS}$, and lepton number violating processes, as discussed above. To get the answer, we assume the seesaw mechanism as 
the origin of the Majorana neutrino masses. In this mechanism the relevant quantity for the leptogenesis is $m_D^{\dagger}m_D$, with $m_D$ being 
the Dirac mass matrix in the basis where the Majorana mass matrix of right-handed neutrinos is diagonalized. The Dirac mass matrix $m_D$ can be expressed 
as $m_D= U \sqrt{D_\nu}R\sqrt{D_R}$, where $U$ is the mixing matrix given in 
Eq.(\ref{unitary}), $D_\nu$ and $D_R$ are the diagonal matrices whose 
eigenvalues are the observable small Majorana masses of left-handed neutrinos and the right-handed Majorana masses, respectively>. $R$ is a complex ``orthogonal" matrix, $R^TR=1$, with 3 independent CP phases, which is otherwise arbitrary. It may be worthwhile to observe $m_D^{\dagger}m_D = \sqrt{D_R}RD_\nu R^\dagger \sqrt{D_R}$. 
That is, the mixing matrix $U$ is not directly related to the leptogenesis. 
We realize that not only the phase in $U_{MNS}$, but also the Majorana phases 
disappear in $m_D^{\dagger} m_D$. 
It, however, will be generally possible to relate the CP violating phases in $U$ to those 
in $R$, responsible for the leptogenesis, once some complementary information, 
such as the absolute values of the elements of $m_D$, is obtained. 
It may also be possible that a relation between the phases of $U$ and those of 
$R$ is naturally realized in a sophisticated concrete model of neutrino mass 
generation.

A road map towards experimental 
verification of leptogenesis may not be easy to draw.
A complete determination of the Majorana mass matrix including the CP
phase may require a variety of lepton number violating reactions such as
$nn \rightarrow e e p p$ (neutrinoless
double beta process),
$e^- \rightarrow \mu^+$,  
$\nu_{\mu} \rightarrow \mu^- \mu^+ \mu^+$, and 
$pp \rightarrow \tau^+ \tau^+ n n$.
Even this much is not sufficient; one further needs a handle to the 
mass scale of the heavy Majorana particle in the seesaw mechanism
to compare with leptogenesis calculation.

We finally make a short comment on processes not discussed so far.
The binding system of stable atoms ($e^-$ + nucleus) has been considered
since in this case a shorter time scale different from the reaction time is
available (like in the case of neutrinoless double beta decay), 
but we did not find a useful system due to the difficulty of
overriding the barrier of nuclear binding.
For instance, the $He$ atom cannot spontaneously decay emitting $e^+$.
As to the decay process,
\(\:
\tau^- \rightarrow \mu^+ + X
\;,
B^+ \rightarrow \mu^+ + \mu^+ + X
\:\)
etc. with $X$ any hadronic state
is conceivable, which has however a branching ratio of order,
\(\:
10^{-4}\,(G_F |m_{\mu \tau}| m_{\tau})^2 \approx 10^{-18}
(|m_{\mu \tau}|/100 eV)^2
\:\)
for $\tau$,
too small even for the tau-charm factory.
(The current upper limit is of order $10^{-6}$.)

Thus, to the best of our knowledge, the processes discussed in the
present work are the best candidates to explore the Majorana nature and
its strength of the neutrino mass matrix, beyond the neutrinoless double
beta decay.
If these tiny rates are experimentally falsified by larger rates,
it definitely implies a new source of lepton number violation besides
the Majorana neutrino mass.

\vspace{1cm}
\begin{center}
{\bf Acknowledgment}
\end{center}

We benefited from discussion with many experimental colleagues.
We would like to thank in particular for 
Y. Kuno, M. Nakahata, I. Nakano, K. Nishikawa, M. Sakuda, R. Tanaka,  
who however are not responsible for the assertions made here.
We also thank K. Hasegawa and K. Tsujimoto for assistance in preparing this manuscript.

\vspace{1cm} 

\end{document}